# Human Mobility in Opportunistic Networks: Characteristics, Models and Prediction Methods

Poria Pirozmand, Guowei Wu, Behrouz Jedari, Feng Xia[*]

School of Software, Dalian University of Technology, Dalian 116620, China
[*]Corresponding author: Feng Xia
E-mail address: f.xia@ieee.org, Tel.: +86 411 87571582

**ABSTRACT**

Opportunistic networks (OppNets) are modern types of intermittently connected networks in which mobile users communicate with each other via their short-range devices to share data among interested observers. In this setting, humans are the main carriers of mobile devices. As such, this mobility can be exploited by retrieving inherent user habits, interests, and social features for the simulation and evaluation of various scenarios. Several research challenges concerning human mobility in OppNets have been explored in the literature recently. In this paper, we present a thorough survey of human mobility issues in three main groups (1) mobility characteristics, (2) mobility models and traces, and (3) mobility prediction techniques. Firstly, spatial, temporal, and connectivity properties of human motion are explored. Secondly, real mobility traces which have been captured using Bluetooth/Wi-Fi technologies or location-based social networks are summarized. Furthermore, simulation-based mobility models are categorized and state-of-the art articles in each category are highlighted. Thirdly, new human mobility prediction techniques which aim to forecast the three aspects of human mobility, i.e.; users' next walks, stay duration and contact opportunities are studied comparatively. To conclude, some major open issues are outlined.

## 1. Introduction

Opportunistic networks (OppNets)(Conti et al., 2010) are emerging paradigms of human-associated ad hoc networks in which mobile users interact with each other based on their geographical proximity. The communication in OppNets is performed in a peer-to-peer fashion using short-range and low-cost mobile devices (such as smartphone and tablet) via Bluetooth or Wi-Fi technologies. In this setting, humans are the main carriers of mobile devices and hence, mobility of devices mirror movement patterns of their owners. This raises the problem of how to generate realistic human mobility traces in order to evaluate the performance of networking protocols in OppNets accurately.

The first generation of networking protocols in traditional mobile ad hoc networks was mainly evaluated using synthetic movement models, such as random way point (RWP)(Bettstetter et al., 2003) or random walk models such as Brownian motion(Groenevelt et al., 2006). However, several research efforts such as (Jungkeun et al., 2003) validate that human mobility is rarely random and random models often fail to analyze the performance of encounter-based protocols in OppNets accurately. However, it should also be noted that human movement and random walks contain some statistical similarities (Injong et al., 2011).

In reality, human mobility is strongly dependent to users' personal and social characteristics and behaviors as well as environmental parameters (Aschenbruck et al., 2011). For instance, mobile carriers which are attracted to specific locations or individuals, may have significant correlation between their respective localization constrains and movement patterns. These different dimensions of human mobility patterns have been characterized in the recent studies. For example, an experimental analysis in (Phithakkitnukoon et al., 2012) demonstrate that users frequently visit the locations with which they have strong social ties. Furthermore, mobile users tend to visit just a few locations, where they spend the majority of their time (Song et al., 2010a). In most cases, they often travel over short distances and rarely migrate long distances (Gonzalez et al., 2008).

Characteristics of human movement can be explored in three main categories: spatial, temporal, and connectivity. The spatial features refer to the trajectory patterns in physical space. Temporal aspects are related to the time-varying features of user mobility, whereas connectivity properties concern the contact information of users. Quite recently, several statistical analysis have been carried out in an effort to better comprehend the properties of human mobility and uncover hidden patterns. Comparatively, it can be seen that there is not a general consensus on the characteristics of human mobility, even on some fundamental features such as the distribution of travel distance. Consequently, it is of paramount importance to obtain insight into these attributes and study the latest findings in this area.

Considering the characteristics of mobility features, an appropriate trace(s) and model(s) for the simulation and evaluation of protocols in OppNets should be wisely selected. As several traces and models that have been proposed are highly similar in nature, it serves a great purpose to have a clear comprehension of the plethora of existing models and data sets. Broadly, human mobility datasets can be obtained in two main methods: realistic and simulation-based models. Majority of the real traces have been registered in bounded environments such as campuses and conferences using Bluetooth or Wi-Fi technologies. In order to generate large scale traces, mobility information has also been acquired using location-based social networks. The simulation-based mobility models are alternative approaches to generate mobility traces synthetically. The main motivation to employ such mathematical methods is to generate scalable and flexible mobility traces. However, the key statistical properties of simulation-based models can be validated using real traces. Despite the fact that the simulation-based models have some spatio-temporal and connectivity dependencies, they can be re-parameterized to be applicable for various scenarios in OppNets.

Human mobility prediction is another challenging issue that has attracted significant attention recently. Undoubtedly, forecasting users' future walks, stay duration and contact properties, based on their mobility characteristics and history has many applications in OppNets. For example, it is remarkably important for an opportunistic data forwarding algorithm to predict the next venue a mobile user will visit, her stay duration and the even the individuals she will contact. By forecasting human mobility, networking protocols can take advantage of the expected information in order to streamline the performance of the algorithms significantly.

In this paper, we study recent solutions for human mobility challenges in OppNets with respect to three major aspects: human movement characteristics, mobility models and prediction methods. Firstly, we categorize the fundamental features of human mobility along the three aspects of spatial, temporal, and connectivity properties. Secondly, commonly used real mobility traces which have been captured using Bluetooth/Wi-Fi technologies and on-line location-based social networking services are summarized. We also present a thorough survey on recently proposed simulation-based human mobility models for OppNets. Thirdly, we categorize human mobility prediction methods into three classes and explore some new techniques in each class. Based on our discussion, we point out some improvements that can be made in the different aspects of human mobility models.

The three topics we study in this paper are closely related to each other. Analyzing different characteristics of human mobility (such as travel distance, contact time) could result in useful indicators and metrics. These measurements are of significant value to uncover meaningful mobility patterns and also to validate available mobility traces. On the other hand, spatio-temporal and contact characteristics of human mobility can be used as useful estimation criteria to predict users' future trajectories.

The remainder of this paper is organized as follows. Section 2 provides an overview on related work. Section 3 covers definitions and terminologies related to characteristics of human mobility. Human mobility models which are categorized into two major classes: trace-based models and simulation-based mobility models are introduced in Section 4. Recently proposed human mobility prediction methods are presented in Section 5. Some major open research issues are presented in Section 6. Section 7 concludes the paper.

## 2. Related Work

There have been some general surveys as well as a few specific surveys for human mobility models. Researchers (Camp et al., 2002) study mobility models for ad hoc networks in two categories: entity models and group models. In addition, they provide a performance evaluation concerning the impact of the different models on multi-hop routing protocols. However, this work mainly focus on random mobility models which are not appropriate movement models for human motion

Researchers (Musolesi and Mascolo, 2009) categorize human mobility models into real world traces and synthetic mobility trace and study advantages and disadvantages of both types. They also, for the first time, introduce the concept of social networks into mobility models. Similarly, (Aschenbruck, Munjal, 2011) provide a survey of real world and simulation-based traces as well as synthetic mobility models for multi-hop wireless networks. The focus of this paper is on mobility traces/models that include position information. Similarly, a categorization on the current mobility models, both from an individual perspective and also from a group perspective is presented in (Andrea and Sofia, 2011). Furthermore, a brief analysis on their applications is presented.

The authors (Karamshuk et al., 2011) present a survey paper on simulation-based mobility models for OppNets. They also discuss properties of movement and extend it with the notion of predictability and patterns. (Munjal et al., 2012) study properties of some synthetic and trace-based mobility models comparatively. Moreover, the changing trends in modeling human mobility are summarized.

Despite the fact that some efforts have been carried out to make a classifiable observation to enumerate mobility issues in OppNets, they have been neither comprehensive nor detailed. To the best of our knowledge, there has never been a clear categorization followed by a comprehensive clarification on human mobility issues in OppNets. To this end, we categorize these issues in three main groups, human mobility characteristics, mobility models, and mobility prediction methods, and explain novel approaches for possible solutions, as outlined in Fig. 1.

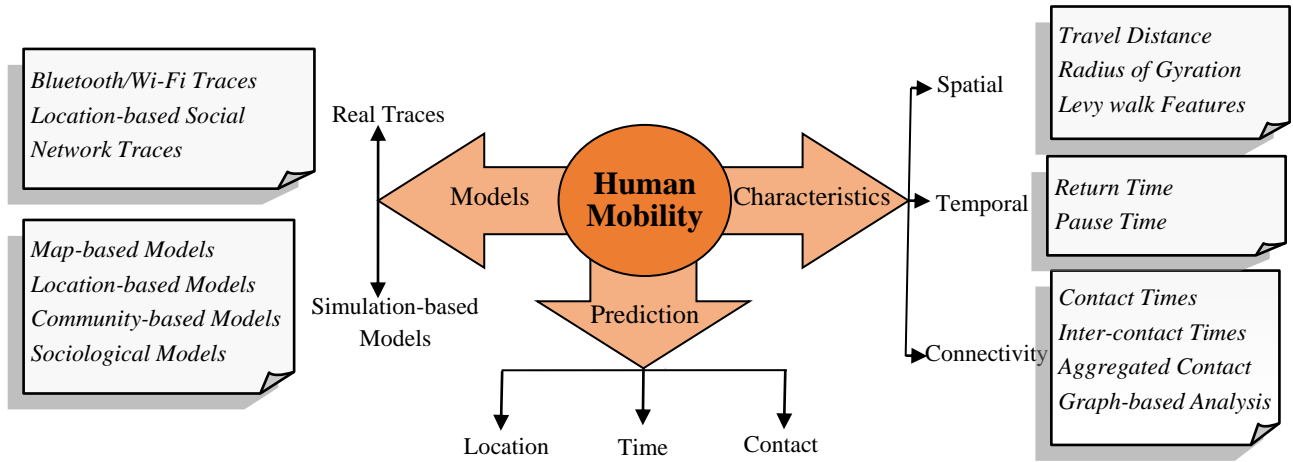

Fig. 1. Three aspects of human mobility namely characteristics, models and prediction methods.

## 3. Human Mobility Characteristics

Statistical features of human mobility are extracted and measured to discover non-homogenous behavior and movement patterns of mobile users in space and time. In addition, they are widely used to predict users' future walks and contacts. These explorations can provide valuable insights into different aspects of human mobility which have many practical applications in wireless networks such as predicting the spread of electronic viruses (Chao and Jiming, 2011), epidemics (Poletto et al., 2013), and recommendation systems (Quercia et al., 2010). In OppNet context, the available characteristics can be exploited to validate the synthetic mobility models (Nguyen and Szymanski, 2012) and study the impact of mobility in the design and performance of routing protocols (Mashhadi et al., 2009, Thakur et al., 2012, Ze and Haiying, 2013).

The most important characteristics of human mobility can be discussed in three main classes, named spatial, temporal and connectivity properties (Karamshuk, Boldrini, 2011). The spatial or geographical features refer to location information of mobile users and their trajectories in physical space. For example, physical length a user travels during a time period. Temporal features explore time-varying properties of human mobility such as the times a user visits some specific locations. Connectivity information of human mobility explore contact and interaction patterns of mobile users which are closely related to social relationships and similarities between mobile users.

In the rest of this section, detailed explanations on human mobility characteristics, based on the recent literature, are investigated comparatively.

### 3.1. Spatial Characteristics

Over the last years, analysing movement of particles in physical space has attracted a lot of attentions by the researchers in the physical, social and geographical sciences. It has been highlighted in almost every quantitative study that a close relationship exists between mobility, space and distance. Random walk models such as Brownian motion (Groenevelt, Altman, 2006) characterize the diffusion of tiny particles with a mean free path (a flight) and a mean pause time between flights. It is shown that the probability that such a particle is at a distance from the initial position after a time has a Gaussian distribution (Gonzalez, Hidalgo, 2008). The mean squared displacement (MSD), which is a measure of the average displacement of a given object from the origin, is proportional to $r$ (normal diffusion). However, there are other objects in the physical world whose mobility cannot be characterized by normal diffusion and its MSD is proportional to $r^\gamma$ ($\gamma<1$) (super-diffusion) (Shlesinger et al., 1982).

Capturing realistic traces from user movements, a huge research effort has been started in order to characterize the spatial properties of human mobility. The first step is to understand how users move across locations. In contract with random dispersal trajectories in particles, human movement on many spatial scales is not random and exhibits high level of spatial regularity (Brockmann et al., 2006, Song et al., 2010b). However, patterns of human movement and random walks contain some statistical similarity (Injong, Minsu, 2011). Furthermore, it is found that users tend to travel most of the time along short distances while only occasionally following very long paths (Gonzalez, Hidalgo, 2008). The origin of this dependence of mobility on space and distance is an open question since a statistically reliable estimate of human dispersal comprising all spatial scales does not exist. In this subsection, the most important spatial properties of human mobility, based on the recent literature, are outlined.

Travel distance or jump size ($\Delta r$) is an important feature of human walks to characterize the spatial dimension of users. Average travel distance can be defined by $\overline{\Delta r(l)} = \sum_{i=2}^{n}|r_i - r_{i-1}|$ where $r_l = \{r_1, r_2, ..., r_n\}$ be the sequence of $n$ geographical displacements user $l$ travels during a time period (e.g., one hour) and $|r_i - r_{i-1}|$ is the distance between locations $r_i$ and $r_{i-1}$. We note that travel distance can range from a few to thousands of kilometers over a periods of time. This measure has a huge impact on the way messages are spread across the network. For example, users that travel over longer distances become bridges between far away communities, and this can be decisive, e.g., for the distribution of messages. However, there are many factors that influence this metric, ranging from means of transportation to job-and-family-imposed restrictions and priorities.

There is an agreement in some literature that the probability distribution of displacements over all users can be well approximated by a truncated power-law (TPL), that is $P(\Delta r) \sim \Delta r^{-(1+\beta)}$ with the displacement exponent or scaling parameter $0 < \beta \leq 2$ (Brockmann, Hufnagel, 2006, Song, Koren, 2010a). This fits the intuition that we usually move over short distances, whereas occasionally we take rather long trips. We note that a power-law is a functional relationship between two quantities, where one quantity varies as a power of another. Furthermore, a truncated power-law distribution follows power-law up to a certain time after which it is truncated by an exponential cut-off. The previous finding are complimented in (Gonzalez, Hidalgo, 2008) with an exponential cutoff, $P(\Delta r) \sim (\Delta r + \Delta r_0)^{-\beta} exp(\Delta r/\kappa)$, where $\beta = 1.75 \pm 0.15$ (mean ± standard deviation), threshold $\Delta r_0 = 1.5$ km, and $\kappa$ a cutoff value varying in different experiments.

The above-mentioned studies suggest the existence of a universal power-law distribution observed for data of humans carrying mobile phones which is in contradiction with observations that displacements strongly depend on where they take place. For instance, a study of hundreds of thousands of cell phones in Los Angeles and New York demonstrate different characteristic trip lengths in the two cities (Isaacman et al., 2010). This observation suggests either the absence of universal patterns in human mobility or the fact that physical distance is not a proper variable to express it. To address this problem, (Noulas et al., 2012) focus on human mobility patterns in a large number of cities across the world upon a dataset collected from a large location-based social network called, Foursquare (https://foursquare.com). Computing the distribution of human displacements in this dataset, it is observed that the distribution is well approximated by a power-law with exponent $\beta =1.50$ and a threshold $\Delta r_0 = 2.87$ which is almost identical to the value of the exponent calculated in (Brockmann, Hufnagel, 2006, Gonzalez, Hidalgo, 2008).

Radius of gyration ($r_g$) is a single scalar quantity which well summarizes the spatial extent of a user's total mobility pattern observed in a time period. According to (Lu et al., 2013), $r_g$ for user $l$ can be defined by $r_g(l) = \sqrt{\frac{1}{n}\sum_{i-1}^{n}|r_j - \bar{r}|^2}$ where $\bar{r} = \frac{1}{n}\sum_{i} r_i$ is the user's center of mass of the trajectory. In comparison to $\overline{\Delta r}$, it can be seen that $r_g$ could be smaller or larger than $\overline{\Delta r}$ in different movement patterns. Traveling in a bounded space results in a small $r_g$ even though the traveler covers a large distance, whereas this value can be larger if the travelers moves with small walks but in a fixed direction.

The $r_g$ probability distribution $P(r_g)$ can be approximated with a truncated power-law, that is $P(r_g)=(r_g+r_g^0)^{-\beta}exp(r_g/\kappa)$ with $r_g^0=5.8km$, $\beta = 1.65 \pm 0.15$ and $\kappa =350km$. This means that the majority of people usually travel in close distances around their home location, while few of them frequently make long travels. To support this result, (Bagrow and Lin, 2012) show that individual mobility is dominated by small groups of frequently visited, dynamically close locations which are called habitats. To support this idea, the reduced $r_g$ is computed for each habitat, considering only those locations and calls contained within that habitat. The population distributions of three habitats show that the spatial extent of habitats tends to be far smaller than the total mobility, often by an order of magnitude, and that most users have a habitat $r_g$ between 1-10$km$.

Some recent work explore human walks to find similar properties between human mobility and Lévy walks. A Lévy walk is a random walk in which the step-lengths have a probability distribution that is heavy-tailed. Intuitively, the Lévy walks consist of many short flights and occasionally long flights where a flight is defined to be a longest straight-line trip of a human from one location to another without a directional change or pause. (Brockmann, Hufnagel, 2006) analyze bank notes and show the Lévy walk patterns in human travels over the scale of a few thousands of kilometers. Similarly, (Gonzalez, Hidalgo, 2008) track location information of mobile phone users at every 2 hours intervals with resolution 2–3 km$^2$ to show that human walks have heavy-tail flight distributions such as Weibull, lognormal, Pareto, and truncated Pareto distributions, and their MSDs are characterized by super-diffusion.

However, both the above-mentioned works have captured coarse-grained location information since their resolution is low (e.g., kilometers) and any flights or travels that occurred between consecutive sampling points cannot be tracked accurately. Consequently, it is hard to apply these features to a detailed simulation in mobile networks, which requires resolutions of a few meters and a few seconds due to short radio ranges of mobile devices. To cope this issue, (Injong, Minsu, 2011) study the mobility patterns of humans up to the scales of meters and seconds and find that human walks at outdoor settings within less than 10km contain statistically similar features as the Lévy walks including heavy-tail flight and the super-diffusion followed by sub-diffusion, which is an indication of heavy-tail flights in a confined area.

3.2. Temporal Characteristics

Analysing trajectories of mobile carriers can help us to discover meaningful temporal patterns in their movements and predict time periods they would stay in some specific locations periodically. During a daily movement, it can be seen that people visit a few number of locations with high probability regularly. The corresponding locations mostly mirror their social roles and characteristics such as their jobs, interests, habits, etc. For example, in a daily routine, a student goes to university, library, gym, etc. at different times of a day and then comes back home at the end of the day.

The probability that a random walker returns regularly to the location he was observed before after $t$ hours is called first-passing time probability (Condamin et al., 2007) and the time period $t$ is called return time. In particular, (Gonzalez, Hidalgo, 2008) show that the probability of finding a user at a location of rank $R$, where $R$ represented the R-most visited location, is approximated by $R^{-1}$, independent of the number of locations visited by the user. Furthermore, it is found that the return probability is characterized by several peaks at 24h, 48h and 72h.

Pause time is another important temporal characteristic which indicates the time period that a user stays in a specific position, i.e., the interval of time when the user's speed is zero or close to zero. Some models prior to social mobility models such as RWP model assumed pause time to be a value randomly picked out from a uniform interval. Analyzing real world data sets, (Brockmann, Hufnagel, 2006, Gonzalez, Hidalgo, 2008) indicate that the probability pause time distribution characterizing human trajectories are fat-tailed, that is $P(\Delta t) \sim (\Delta t)^{-(1+\beta)}$ with $0 < \beta \le 1$, where $\Delta t$ is the time spent by an individual at the same location. For example, (Song, Koren, 2010a) find that $P(\Delta t)$ follows $P(\Delta t) \sim |\Delta t|^{-(1+\beta)}$ with $\beta = 0.8 \pm 0.1$ and a cutoff of $\Delta t = 17$ hours, probably capturing the typical awake period of an individual.

Self-similar least action walk (SLAW) (Kyunghan et al., 2009) is one of the first models to consider pause time. Based on a global notion, the authors define pauses considering a TPL distribution, where each visit point has a specific pause value, and the average pause time is adjusted for the whole trip to be completed in 12 hours. However, the way the individual routine was modeled is still artificial in the sense that it just takes into account a potential split in terms of time, and disregards any time and space correlation. Quite recently, (Ribeiro et al., 2012) propose pause time modelling assuming that there are several social attractiveness properties instead of providing an approximation to traces. Hence, a function is proposed that takes into consideration the attractiveness between nodes to model pause times. As experimental basis, one of the most popular mobility models, CMM is considered for the modeling, which provides a movement function based upon a social attractiveness function.

3.3. Connectivity Characteristics

In OppNets, a contact occurs when two mobile devices are within mutual radio transmission range of each other. Since every contact is an opportunity to forward content and bring it probabilistically closer to a destination (or a set of destinations), understanding statistical properties of contacts is vital for the design of algorithms and protocols in intermittently connected networks. Contact times (CTs) or contact durations is defined as the time intervals during which two devices are in a radio range of one another. Depending on the scenario considered, it can be the duration either of a Bluetooth association or of the staying under the coverage of the same access point. The CTs is a main factor in determining the transmission capacity between encounter devices in a mobile network. Obviously, two devices can exchange more messages with longer CTs.

Inter-contact times (ICTs) is another important factor which is essential building block for the network properties. An ICT is the amount of time elapsed between two successive contact periods for a given pair of devices. It characterizes the frequency with which data can be transferred between networked devices. This indicator is particularly important in OppNets, since it defines the frequency and the probability of being in contact with the recipient of a message or a potential message carrier in a given time period. However, some new metrics like inter-transfer time (Chul-Ho et al., 2013) have been also introduced which aim to characterize and evaluate the actual off-duration of the connectivity link in OppNets.

Characterizing the ICTs is essential for analysing the performance of networking protocols in OppNets. For example, shorter ICTs between two nodes means that the nodes encounter each other often and they could exchange data among each other directly whereas for longer ICTs, other encounter nodes in the network could be selected as data carrier to deliver data. It can also be deduced that nodes with shorter ICTs do not have new information to share every time they meet. On the other hand, new data is shared among nodes with longer ICTs.

On an individual pair-wise level, most trace analysis research has focused on CTs and ICTs statistics to investigate whether these distributions are power-law, have exponential tail, have an exponential cut-off, or have qualitatively different behavior from on pair to another[1]. The first body of work, to the best of our knowledge, which highlights the importance of ICTs for characterizing OppNets is presented in (Chaintreau et al., 2007) using eight distinct experimental real-world data sets. They also find very important theoretical results showing that naive forwarding protocols may diverge in homogeneous networks if individual pair ICTs are heavy tailed. Analyzing the same datasets, these results are refined in (Karagiannis et al., 2010) that the distribution of ICT does not only follow a power-law, but exhibits an exponential cutoff.

Researchers (Conan et al., 2007) explore to uncover the distributions of individual pair ICTs. Based on extensive analysis, they show that this feature is actually heterogeneous, and that an exponential distribution fits well a significant fraction of individual pair ICTs, while Pareto and Lognormal distributions also show a good fit with other subsets of the pairs. The authors (Gao et al., 2009) also analyze the MIT Reality Mining trace (Eagle and Pentland, 2006), finding that exponential distributions fit over 85 percent of the individual pair ICTs.

---

[1] We note that in a network of n nodes, there are n(n − 1)/2 inter-contacts distributions.

Studying the characteristics of the ICT on pair-wise level is difficult since there are many such distributions for each individual, and that some of them may only include a few observed values. To tackle this problem, some of the literature use aggregate ICTs distribution, i.e. the distribution obtained by considering the samples from all pairs together, to characterize the properties of contacts between mobile users. There is an agreement in some literature such as (Boldrini and Passarella, 2010, Injong, Minsu, 2011, Kyunghan, Seongik, 2009) that aggregate ICTs distributions feature a power-law with exponential cutoff, and do not pay attention to the possible difference of the individual level distributions. Researchers (Chaintreau, Pan, 2007) also analyze a popular set of real traces finding that the aggregate ICTs can be approximated with a Pareto distribution with shape less than one. This result is then softened in (Karagiannis, Le Boudec, 2010), who reanalyze the same traces and verify that the aggregate ICTs distribution might indeed present an exponential cutoff in the tail, following the Pareto shape.

With respect to the above works, (Passarella and Conti, 2013) provide an analysis of the dependence and the key differences between individual pairs and aggregate ICTs distributions. Based on detailed analysis, it is shown that in heterogeneous networks, when not all pairs contact patterns are the same, most of the time the aggregate distribution is not representative of the individual pairs distributions, and that looking at the aggregate can lead to completely wrong conclusions on the key properties of the network.

Some proposals focus on the graph structure of human contacts because of the social nature of human mobility. In this trend, two main steps have been conducted. First, contacts are presented in a compact way as a directed or weighted contact graph, where the directions show contact order and weights express contact frequency or contact duration between two encounter nodes. In some cases, a social graph is constructed which is an intuitive source for many social metrics. In a social graph, nodes correspond to social entities (e.g. individuals), and edges represent social ties between the social entities that may be inferred from the frequency of observed contacts, shared interests, or geographic preferences. In the second step, social network analysis (SNA) techniques (e.g., connectivity metrics, community detection, etc.) can be utilized to explore and reveal meaningful contact patterns in the underlying mobility scenario.

Considering the above descriptions, some recent work investigate contact properties of human mobility based on the graph structure and social network analysis methods. Researchers (Shunsen et al., 2010) present a social-based human mobility model based on centrality and community structure. Instead of extracting communities from artificially generated social graphs, their model manages to construct an overlapping community structure which satisfies the common statistical features observed from distinct real social networks. Similarly, physical proximity and overlapping community concepts have been utilized in (Yoneki et al., 2009) in order to analysis dynamics of meeting times and inter-meeting times between mobile users and infer meeting groups.

The authors (Hossmann et al., 2011) represent real world mobility traces as a weighted contact graph to show that the structure of human mobility has small-world properties. Furthermore, a community detection mechanism is utilized to show that human mobility is strongly modular. Furthermore, (Mayer and Waldhorst, 2011) utilize spectral graph theory to analyze the impact of the underlying graph characteristics on ICTs. Based on synthetically and real mobility traces, it is found that ICTs are strongly influenced by graph structure in random and social mobility models. Secondly, real-world city maps do not exhibit sufficient difference in structure to effectively influence ICTs of a social mobility model.

Taking the above discussions into account, it can be seen that various aspects of movement and mobility patterns have been characterized in the recent literature in different ways. Despite the fact that human motion exhibits structural patterns due to geographic and social constraints, there is not a general agreement on most of their specifications and properties even on some basic features such as travel or ICTs distributions. In some cases, the main reason is that reliable large scale human mobility data has been hard to acquire which makes this research area quite challenging and there exists several open issues without any proper answer.

## 4. Human Mobility Models

Availability of human movement traces which are collected from real-life human mobility or generated using simulation-based methods make it possible to analysis and explore pattern of trajectories of mobile carriers in OppNets. Traces from the real world are mostly recorded from opportunistic contacts between the users using small portable wireless devices in campuses, conferences, entertainment environments, etc. However, most of these realistic movement traces are not scalable, flexible or accessible for the public. To this end, synthetic models have been proposed to capture the movement patterns of nodes in a realistic way. In the rest of this section, state-of-the art of human mobility, based on the recent literature, are categorized and studied. Furthermore, the most important properties are featured.

### 4.1. Real Mobility Traces

In the last few years, several real life human mobility traces have been collected in order to explore human motion and evaluate the performance of human-associated networking protocols. The mobility traces have been acquired using various kinds of communication systems and devices such as a global positioning system (GPS), user phone calls, and WLAN Access Point associations. Because of their realistic nature, the obtained traces provide huge volume of mobility information in a large scale, although they are captured on different accuracy and granularities. Depending upon collection devices and filtration techniques, they include different features of human mobility such as location, time and contact information.

Quite recently, the performance of networking protocols and algorithms in infrastructure-less settings like OppNets have been evaluated using realistic and non-random mobility datasets. In contrast to traditional ad hoc networks, users' movements in this setting have some regularities. Therefore, their spatio-temporal, context and contact patterns can be captured aim to discover new properties in human mobility and streamline networking protocols in OppNets.

Broadly, real world human mobility traces for OppNets are obtained using two different methods: Bluetooth/Wi-Fi enabled devices and location-based social networking services. The first category of traces can be collected directly or indirectly using Bluetooth or WLAN ad hoc wireless technologies. These methods are able to collect users' location and contact information in medium-scale environments such as conferences, universities, etc. Commonly, two types of mobile devices are carried by mobile participants to collect mobility data in this method. There are particular participants those carry special sensor devices called internal devices. Those which carry Bluetooth/Wi-Fi enabled devices are called external devices. However, capturing users' contact traces using short range wireless sensors is usually complex and expensive. Furthermore, contact information obtained in this method are limited with respect to the practical number of sensors that can be deployed as well as the number of available human volunteers.

Table 1 outlines common real world mobility datasets and provides brief descriptions for them. Furthermore, their important characteristics are featured in this table. We list traces that are collected via Bluetooth or Wi-Fi technologies and they are publically available. Most of these data can be accessed on the CRAWDAD (crawdad.cs.dartmouth.edu) archive.

Location-based social networks (LBSNs) are alternative source of human mobility traces which are collected from online services. In this method, users share their location information with their friends by checking-in their visiting locations and putting messages, or other location-related information on social networks like Facebook. In addition, their context, social ties and offline information can also be linked to thir mobility traces. Comparing to the Bluetooth/Wi-Fi data, mobility traces which are collected in this method are reliable and scalable which allow researchers to access to a large volum of accurate mobilty data. However, some incentive methods should be utilizeed in these methods to encourage users to report their visiting places persistencly.

Gowalla (http://blog.gowalla.com/) is a pioneering LBSN service created in 2008 which allows users to report their location information in a website or using a mobile application and share their locations with their friends. Users in this system could broadcast their location information to their friends. Gowalla users can also connect their accounts to their Twitter account. Gowalla application programming interface (API) allow researchers to access to their contents such as users' check-ins, social relationships and friend information. The available data are of great value to evaluate human mobility dynamics in OppNets. However, Gowalla has been purchased by Facebook recently and is no longer accessible freely.

Recently, several researches have been conducted to analysis properties of human mobility using Gowalla data. The authors (Nguyen and Szymanski, 2012) study the power and limitation of the data captured from Gowalla for providing insights on how distance limits the possibility of friendship. Similarly, (Allamanis et al., 2012) presents social links created and places visited by users in Gowalla data as an undirected graph in order to study the evolution of the social graph of a location-based service and the effect of spatial and temporal factors on the growth of the network.

Stumbl is a Facebook API that allows users to create an application and access to location, communication and content information of Facebook users. In a recent work (Hossmann et al., 2012), the mobility, social and communication information of mobile users are explored using Gowalla and Stumbl data. Based on the evaluation, it is concluded that the three dimensions of tie strength, i.e., mobility, social ties, and communication depend on each other. For instance, it can be seen that social ties and mobility ties as well as communication and mobility dimensions are tightly related to each other.

Foursquare (https://foursquare.com) is a leading LBSN application with more than 20 million users as of April 2012. The Foursquare mobile application allows users to check in venues using a website or via their smartphones. Submitting a venue, a user is asked to provide a few attributes of the venue such as venue's name, address, location, zip code, and etc. Foursquare API can return a list of venues in a region which can be specified by the latitudes and longitudes of the region bounding box. The authors (Yanhua et al., 2013) study the common characteristics of popular venues in Foursquare. Specifically, they investigate how the completeness of venue profile information impacts the venue popularity and conclude that venues with more complete profile information are more likely to be popular. Similarly, (Noulas et al., 2011) analysis the geographical and temporal dynamics of collective user activity on Foursquare and show how check-ins information can be utilized to uncover human daily and weekly patterns, urban neighborhood properties and recurrent transitions between different activities.

**Table 1**
Characteristics of common real-life Bluetooth/Wi-Fi connectivity traces. (Patterns: "NA" – stands for not available cases. In the column Network Type, "B" – stands for Bluetooth and "W" – stands for Wi-Fi).

| Trace | Characteristic | Network type | Duration (day) | No. internal device | No. external device | No. internal contacts | No. external contacts |
|---|---|---|---|---|---|---|---|
| UCSD (McNett and Voelker, 2005) | In the UCSD, approximately 300 wireless PDAs running Windows CE were used to collect Wi-Fi access point information periodically for 11 weeks. | W | 77 | 275 | NA | 195364 | NA |
| Proximity | This data includes a number of traces sightings by groups of users carrying small devices in office environments, conference environments, and city environments. (http://kdl.cs.umass.edu/data/msn/msn-info.html) | B | NA | NA | 27 | NA | NA |
| Dartmouth (Henderson et al., 2004) | This dataset about traffic in the access points was extracted from the SNMP logs of the Dartmouth College Campus between 2001 and 2004. | W | 114 | 6648 | NA | 4058284 | NA |
| Infocom 2005 (Hui et al., 2005) | This trace includes sightings by groups of users carrying iMotes in Conference IEEE Infocom in Miami. | B | 4 | 41 | 264 | 22,459 | 1,173 |
| Infocom 2006 (Chaintreau et al., 2006) | This trace includes sightings by groups of users carrying iMotes in Barcelona, Spain. | B | 4 | 98 | 14,036 | 191,336 | 63,244 |
| MIT Reality Mining (Eagle and Pentland, 2006) | This project equipped students and staff at MIT for the entire 2004–2005 academic year. | B | 246 | 89 | NA | 114046 | NA |
| Toronto (Su et al., 2006) | A trace of Bluetooth activity in different urban environments to determine the feasibility of a worm infection. | B | 16 | 23 | NA | 2802 | NA |
| Intel (Chaintreau, Pan, 2006) | This trace includes sightings by groups of users carrying iMotes in Intel Research Cambridge Corporate Laboratory. | B | 6 | 8 | 92 | 1091 | 1173 |
| ETH (Tuduce and Gross, 2005) | Using the same methodology as for Dartmouth, a trace collected at the ETH campus. | B | 75 | NA | 285 | NA | NA |
| Ile Sans Fils (Scellato et al., 2011) | This dataset was collected from a large number of hotspots which were deployed mostly in cafes, restaurants and bars, libraries in Canada. | W | 1095 | 140 | 45000 | NA | NA |
| Roller (Tournoux et al., 2009) | Contains the contacts from a set of people who participated in a rollerblading tour in Paris. | B | 3 h | NA | 62 | NA | NA |
| SIGCOMM (Pietilanen and Diot, 2012) | The dataset is collected during SIGCOMM 2009 conference. Each device is initialized with the participants Facebook profile and the list of friends. | B | 150 | NA | 100 | NA | NA |
| Nottingham/mall (Galati and Greenhalgh, 2010) | The experiment was conducted on a shopping mall from 9am till 9pm, which has a surface area of 10,880 square meters. | B | 6 | 2 | 23 | 284492 | 60223 |
| MDC (Laurila et al., 2012) | In MDC, smart phones that collect behavioral data were allocated from Lake Geneva region. | B | 365 | 200 | 39 | NA | NA |
| uiuc/uim (Long et al., 2011) | This is the dataset collected by the University of Illinois Movement framework using Google Android phones. | B/W | 150 | NA | 123 | NA | NA |
| Europe (Andrea and Sofia, 2011) | Collected by the University of Illinois Movement (UIM) framework using Google Android phones. | B | 180 | NA | 100000 | NA | NA |
| upb/mobility (Ciobanu et al., 2012) | This is the data collected from Android phones in University Politehnica of Bucharest. | B | 35 | 22 | 655 | 341 | 1127 |
| UIM Trace (Vu et al., 2011) | This dataset was collected by the University of Illinois using Google Android phones. | B/W | 19 | 16 | 50 | 76000 | NA |

## 4.2. Simulation-based Mobility Models

Simulation-based mobility models aim to mimic movement of humans in a real life and simulate their mobility patterns using parametric methods synthetically. These models give the opportunity to evaluate networking protocols in different scenarios, and test their robustness to different mobility behaviors. Different kinds of rules can be defined in these models to make the mobile nodes follow a popularity distribution when selecting the next destination individually, or moving as a group.

Due to several reasons, simulation-based mobility models have been largely preferred for the evaluation of human-associated networking protocols. Firstly, majority of real traces are environment specific, i.e., they are collected in universities or conferences, and are not scalable. Secondly, they are not controllable and flexible for changing system parameters such as node density and node velocity. In addition, the publicly available traces are limited. These problems forced researchers to use simulation-based models, where the parameters of the mobility models can be modified according to problem specifications. However, characteristics of these models should be validated using real world traces such a method in (Nguyen and Szymanski, 2012).

Broadly, simulation-based mobility models can be categorized into four main classes, namely map-based models, location-based models, community-based models, and social-based models. In the rest of this subsection, pioneering mobility models in each category are introduced.

### 4.2.1. Map-based Models

The map-driven mobility models extract movement features of real world traces in order to reproduce scalable mobility traces using simulation methods synthetically. Working day movement (WDM) (Ekman et al., 2008) is a pioneering map-based method that is able to produce CTs and ICTs distributions that follow closely the ones found in the traces from the real-world measurement experiments. This model incorporates some sense of hierarchy and distinguishes between inter-building and intra-building movements.

Agenda driven mobility model (ADMM)(Zheng et al., 2010) is another well-known model in this class. ADMM utilizes national household travel survey (NHTS) information from the U.S. Department of Transportation to obtain activity and dwell time distributions. In this work, a mobile ad hoc network in an urban scenario is simulated in order to analyze the geographic features of the network topology. The impact of the model on routing performance is also investigated in this work.

SAME (Xuan et al., 2012) is a mobility model of daily activities which is based on the analysis and conclusion of students' habits and customs in campus environments. Mobility in SAME is divided into five sub-models, including dormitory sub-model, learning sub-model, eating sub-model, out activity sub-model and transport sub-model which describe the moving instruments between former sub-models. Simulation results turn out that comparing to the WDM, SAME is much closer to actual trace statistics.

### 4.2.2. Location-based Models

The class of location-based mobility models aims to represent user mobility patterns using a set of preferred locations. Various attributes and relationships can be identified between the preferred locations and users in these models which determine movement trajectories of the users across these locations. Sociological orbit aware location approximation and routing (SOLAR) (Ghosh et al., 2005) is one of the first proposals in this category. The SOLAR is a mobility framework which takes advantage of the macro-mobility information obtained from the sociological movement pattern of mobile users. This model is motivated by the observation that the mobility of a mobile user exhibits a partially repetitive orbital pattern. Although the SOLAR is not general enough to be realistic in conventional ad hoc networks, it can be specifically used without a need for constant location updates and flooding that makes it suitable to OppNets.

Time-variant community model (TVCM) (Wei-jen et al., 2007) is a prominent movement model in this class to capture the important mobility properties observed from daily lives. In this method, some locations called communities are defined to be visited by each node in order to capture skewed location visiting preferences. In addition, time periods with different mobility parameters are used to create periodical re-appearance of nodes at the same location. This approach is extended in (Kyunghan, Seongik, 2009) by proposing the self-similar least action walk (SLAW) which is one of the first mobility models to reproduce the preferences for shorter trips. A process is called self-similar if the aggregated processes (i.e., the processes obtained by averaging the original process over non-overlapping blocks) are highly correlated. The model matches the ICTs distribution of the real traces and is also able to model the pause time. The performance evaluation analysis of SLAW generated traces shows that this method demonstrates social contexts present among people sharing common interests or those in a single community such as university campus, companies and theme parks.

Small world in motion (SWIM) (Mei and Stefa, 2009) is another prominent mobility model for ad-hoc networking approach based on location preference. SWIM is relatively simple which is easily tuned by setting a few parameters. In this model, a randomly and uniformly chosen point over the network area is assigned to each node. The node then selects the destination points of their movement based on their popularity among all nodes and their distance from the home point.

In a recent work, (Nguyen et al., 2011) propose spatio-temporal parametric stepping (STEPS), a simple parametric mobility model which can cover a large spectrum of human mobility patterns. STEPS makes abstraction of spatio-temporal preferences in human mobility by using a power law to rule the nodes movement.

4.2.3. Community-based Models

A community is often defined as a group of network members with stronger ties to members within the group than to members outside the group. The organization of users in community-based mobility models leads to spatial and social dependencies among the users. In this paradigm, if two nodes belong to the same community, they tend to spend more time together and follow their movements when they transit to a new location. The first example of this class of models is presented in (Herrmann, 2003). In this model, users are organized in several communities, and then each community is associated to a physical location. The movements of users follow a predefined schedule across the locations associated with their group. This work, however, lacks a rigorous mathematical definition of the relationships among users.

The community-based mobility model (CMM) (Musolesi and Mascolo, 2007) is a flexible model which is directly driven by a social network. In CMM, nodes belong to a community are called friends, while nodes in different communities are called non-friends. At the beginning, the movement area is divided into some regions as a grid and each community is assigned into a cell of the grid. A link is established between all the friend and non-friend nodes in the network which will be used later to drive node movements. In this model, nodes move between the communities based on node attraction feature. In other words, nodes of a community follow the movements of the first node of that community that has decided to exit the physical location. For example, when a node with high attraction value decides to travel to another community, all nodes belong to the same community follow the movements of the node.

The authors (Boldrini and Passarella, 2010) highlight the shortcoming of CMM which is called gregarious behavior of nodes: *"all the nodes of a community follow the first node that has decided to exit the community"*. To tackle this issue, home-cell community-based mobility model (HCMM) is proposed which considers both node and location attraction. HCMM maintains the social model of CMM, but some nodes have also social links with communities other than the home which is called foreign community. In this model, each node is initially associated with a specific community, and has social ties with all the other members of its home community. Some special nodes also have social links with foreign communities other than the home community. A node assigned into its home community moves towards a given community with a probability proportional to the number of ties with nodes of that community. Simulation results indicate that CTs and ICTs distributions for HCMM match that of CMM model, which has the same pattern of the traces such as the Cambridge, UCSD, and Infocom 2005 real world traces.

Enhance community mobility model (ECMM) (Vastardis and Yang, 2012) is an extended version of CMM and HCMM models which follows preceding community-based approaches. The main contribution in this model is the introduction of new features, such as pause periods and group mobility encouragement which have not been considered in the previous community-based models. Additionally, ECMM enables researchers to arbitrarily select a social model as the trace generation process input, while at the same time generates traces with high conformance to that social network.

Geo-community (Geo-CoMM) (Zignani, 2012) is another community-based model which reproduces the spatio-temporal, and social features of real mobility datasets. The model is based on the quantities that guide human mobility and their probability distributions by directly extracting their setting from the statistical analysis of GPS-based traces. In Geo-CoMM, users move within a set of geo-communities, i.e. locations loosely shared among people following speed, pause time and choice rules whose distribution is obtained by statistical analysis methods.

4.2.4. Sociological Models

Social characteristics, behaviors and interaction patterns of users in a real life can be exploited to design the basic mechanisms of human movement. Sociological mobility models can be considered as the application of social network theory on the field of mobility modeling to formalize social interactions as the main driver of human movements.

Sociological interaction mobility for population simulation (SIMPS) (Borrel et al., 2009) is a prominent social-based model in which mobile entity move according to two behavioral rules: social interaction level, i.e. the personal status; and social interaction needs, i.e. the social needs for individuals to make acquaintances. These two behaviors alternate according to a feedback decision-making process which balance the volume of current social interactions against the volume of interactions needed by the node. The main drawback of SIMPS is that temporal and spatial regularities such as pause time are not considered in this model.

GeSoMo (Fischer et al., 2010) is a social-based mobility model which separates the core mobility model from the structural description of the social network underlying the simulation. GeSoMo receives a social network as input and creates a mobility trace which is a schedule for the movement of each individual node in the input social network such that this trace creates meetings between the nodes according to their social relations. In this model, the attraction between nodes is defined based on node attraction, location attraction and node repulsion (i.e. negative attraction).

**Table 2**
Comparison of simulation-based human mobility models. (Patterns: "√" – if the model satisfies the property, "×" if not, and "–" for ambiguous cases.)

| | Model | Characteristic | Spatial | | | | | Temporal | | Connectivity | | | | |
|---|---|---|---|---|---|---|---|---|---|---|---|---|---|---|
| | | | Travel distances | Radius of gyration | Community | Attraction Node | Attraction Location | Return time | Pause time | Contact times | Inter-contact times | Aggregate ICTs | Contact graph | Social graph |
| **Map-based models** | WDM (Ekman, Keranen, 2008) | A combination of different movement models that is able to produce inter-contact time and contact time distributions. | √ | × | × | × | × | √ | √ | √ | √ | × | √ | × |
| | ADM (Zheng, Hong, 2010) | Contains personal agenda, geographic map, and motion generator components that model social activities, geographic locations, and movements of mobile users. | √ | – | × | × | × | × | √ | – | – | – | × | × |
| | SAME (Xuan, Yuebin, 2012) | A model of students' daily activities based on the analysis and conclusion of students' habits and customs in campus environments. | – | – | √ | × | × | – | √ | √ | √ | × | √ | × |
| **Location-based models** | SOLAR (Ghosh, Philip, 2005) | Each node selects a subset of predefined sets of locations and moves between them based on a customizable behavior. | × | √ | √ | × | × | – | √ | – | – | × | × | × |
| | TVCM (Wei-jen, Spyropoulos, 2007) | Model the spatio-temporal preferences of human mobility by creating community zones. | × | × | √ | × | √ | – | √ | – | – | – | √ | × |
| | SLAW (Kyunghan, Seongik, 2009) | One of the first mobility models to model pause time. Based on a global pause time perception, pause time is then randomly defined for each individual node. | √ | √ | √ | × | × | – | √ | √ | √ | √ | √ | × |
| | SWIM (Mei and Stefa, 2009) | Nodes select the destination points of their movement based on their popularity among all nodes and their distance from the home point. | √ | √ | √ | × | × | × | √ | √ | √ | × | × | √ |
| | STEPS (Nguyen, Sénac, 2011) | Makes abstraction of spatio-temporal preferences in human mobility by using a power law to rule the nodes movement. | √ | – | √ | × | √ | – | √ | √ | √ | √ | × | √ |
| **Community-based models** | CMM (Musolesi and Mascolo, 2007) | Nodes are assigned to a number of subareas using preferential attachment. The attractiveness of one area is determined by the current number of nodes assigned to that area. | × | – | √ | √ | × | – | – | √ | √ | × | × | √ |
| | HCMM (Boldrini and Passarella, 2010) | Combines notions about the sociality of users with spatial properties observed in real users movement patterns. | √ | – | √ | √ | √ | × | × | √ | √ | × | × | √ |
| | ECMM (Vastardis and Yang, 2012) | Follows approaches in (Musolesi and Mascolo, 2007) and (Boldrini and Passarella, 2010), but introduces new features, such as pause periods and group mobility encouragement. | × | – | √ | √ | √ | × | √ | √ | √ | × | × | √ |
| | Geo-CoMM (Zignani, 2012) | In this model, people move within a set of geo-communities, following speed, pause time and choice rules whose distribution is obtained by the statistical analysis. | √ | × | √ | × | × | √ | √ | √ | √ | √ | √ | × |
| **Sociological models** | SIMPS (Borrel, Legendre, 2009) | Derives the motion of users in a way that individuals' movements are governed by both their social relationships and geographically surrounding individuals. | × | × | × | √ | × | √ | × | √ | √ | × | √ | × |
| | GeSoMo (Fischer, Herrmann, 2010) | Separates the core mobility model from the structural description of the social network underlying the simulation. | √ | – | × | √ | √ | √ | × | – | √ | √ | × | √ |
| | SPoT (Karamshuk et al., 2013) | A mobility framework that takes a social graph as input. Then, the spatial and temporal dimensions of mobility are added. | × | × | √ | × | × | – | √ | √ | √ | √ | × | √ |

12Social, sPatial, and Temporal mobility framework (SPoT) (Karamshuk, Boldrini, 2013) considers the three main aspects of mobility to provide a flexible and controllable mobility framework rather than a model. SPoT aims to generate different mobility models by modifying mobility properties (e.g., ICTs) simultaneously. To handle such a model, the model takes the social graph as an input. Then, the spatial dimension is added by generating an arrival network. Based on the input social graph, communities are detected and mapped into different locations. In the next step, users belonging to the same community share a common location where the members of the community meet. Then, users visit these locations over time based on a configurable stochastic process. Furthermore, real traces from some LBSNs such as Gowalla and Foursquare are utilized in this work to characterize the temporal patterns of user visits to locations. Based on the analysis, it is shown that a Bernoulli process approximates user visits to locations in the majority of cases. Table 2 summarizes the main characteristics of the simulation-based mobility models.

## 5. Human Mobility Prediction

Tracking users' future movement behavior and quantifying their predictable regularity have been topics of considerable interest in recent years. Human mobility prediction capabilities can be used in various exciting mobile applications such as data sharing, epidemic modeling, traffic planning, and disaster response. For example, human mobility can be used for predicting the spread of electronic viruses and malwares (Chao and Jiming, 2011). Contact prediction between mobile carries has also been widely utilized in some proposals such as (Pan and Crowcroft, 2008, Quan et al., 2012) aim to streamline routing and data forwarding protocols in OppNets. As an example, if a student has a message to send to her classmate, she knows that she will have a high probability to meet the receiver during school time and at the school.

Given the characteristics of users' mobility patterns, predicting next locations, stay durations and future contacts of individuals or a population are crucial issues. There are many factors which influence a user's future mobility patterns ranging from their personal and social factors to parameters of the environment. On the other hand, characteristics of the location data set can affect the forecasting accuracy significantly. Recently, several human mobility prediction methods have been proposed to demonstrate that human mobility can be predicted to a greater extent. For instance, in an experiment on more than 6000 users on the Dartmouth dataset, it is found that the best location predictor has an accuracy of about 65–72% (Song et al., 2006). Similarly, two groups of leading network scientists in (Lu, Wetter, 2013, Song, Qu, 2010b) find that human behavior is 93% and 95% predictable, respectively.

In the last few years, extensive researches have been carried out to forecast different aspects of human mobility in OppNets. The existing human mobility prediction methods can be discussed in three groups. The first category of the proposals have strived to predict the next visiting locations of mobile users based on history of their trajectories. In some recent work, it has been aimed to predict a user's future arrival time to a corresponding location and his/her stay duration. Finally, some methods have strived to predict contact probability of mobile users (i.e., who will contact each other?). We note that the proposals those predict all future aspects of human mobility, (i.e., location, time, and contact properties) are discussed in the third group. In the rest of this section, some new prediction techniques in each category are presented.

5.1. Location Prediction

Predicting users' future visiting locations has many applications in pervasive computing applications such as location-based recommendation and advertisement dissemination systems. Hence, mobility traces of users have been extensively analyzed in order to gain insight about humans' mobility patterns and forecast their future locations accurately.

Analyzing trajectories of 6000 WLAN users, (Song, Kotz, 2006) compare four major families of domain-independent location predictors using the Dartmouth dataset. Based on the comparative implementation, it is found that the low-order Markov predictors performed as well or better than the more complex and more space-consuming compression-based predictors. The authors also concluded that Markov predictors based on *n* previous locations ($n \geq 3$) are less precise.

Considering the above-mentioned results, several Markov chain based methods have been presented in order to predict a user's future locations. A mixed Markov chain model (MMM) is proposed (Asahara et al., 2011) to predict a pedestrian's next location in three steps. In the first step, a statistical model is used to extract pedestrians' past trajectories. Then, a new user's tracking data is collected. Finally, the user's next position is predicted by using the statistical model and the user's tracking data. The experimental analyze in a shopping mall demonstrates that the highest prediction accuracy of the MMM is 74.4%. Similarly, (Gambs et al., 2012) propose a prediction algorithm called *n-MMC*, based on mobility Markov chain (MMC) (Gambs et al., 2010), that keeps track of the *n* previous visited locations and predict a user's next location. The evaluation of three mobility datasets demonstrates an accuracy for the prediction of the next location in the range of 70% to 95% as soon as n = 2.

Quite recently, (Lu, Wetter, 2013) analyze the travel patterns of 500,000 individuals using mobile phone data records and study a series of Markov chain based models to predict the actual locations visited by each user. By measuring the uncertainties of movements using entropy, it is found that the theoretical maximum predictability is as high as 88%. To verify whether such a theoretical limit can be approached. Implementation results in this paper show that M chain models can produce a prediction accuracy of 87% for stationary trajectories and 95% for non-stationary trajectories.

Researchers (Daqiang et al., 2013) explore real phone traces and find that there is a strong correlation between the calling patterns and co-cell patterns of mobile users. Based on this idea, a location perdition algorithm called NextCell is proposed that aims to enhance the location prediction by harnessing the social interplay revealed in cellular call records. The performance of this method is evaluated using the MIT Reality Mining trace. Experimental results show that NextCell achieves higher precision and recall than the state-of-the art schemes at cell tower level in the forthcoming one to six hours.

All the above-mentioned approaches assume sufficient sets of training data to predict next locations of mobile users. However, in some cases, this data is not available for new users which leads to lower prediction accuracy. To tackle this issue, (McInerney et al., 2012) measure similarities between new and existing users. Then, a hierarchical Bayesian model for matching the locations of a new user with those of existing users is proposed to enhance the accuracy of the prediction. Evaluate this framework using real life location habits of 38 users shows that accuracy on predicting the next location of new users is improved by 16%, comparing to the benchmark methods.

Majority of the explored methods in this subsection consider users' spatial and temporal features separately to predict their future locations. Integrating both spatial and temporal information for location prediction is challenging since temporal aspect of human mobility includes considerable uncertainty compared to spatial movements. Taking both the features into consideration, (Gao et al., 2012) proposed a location prediction model by applying smoothing techniques to capture the spatio-temporal context of user visits. Similarly, (Huang et al., 2012) consider movement behaviors of users in space and time, and proposed a prediction framework for semantic place prediction. The core idea of this proposal is to extract 54 features in the MDC data set in order to represent end users' behaviors in each place related to its semantic.

5.2. Time Prediction

Some human mobility prediction methods have studied temporal forecasting of human mobility. However, uncovering the temporal aspects of human mobility is challenging because humans' temporal behavior includes considerable uncertainty compared to spatial movements. Majority of the methods in this category have aimed to predict a user's arrival time to a location and his/her staying duration. In this subsection, some representative proposals about temporal prediction of human mobility are discussed.

The authors (Do and Gatica-Perez, 2012) address the prediction of user mobility when they arrive to or leave a place. In other words, they strive to predict a user's mobility considering two questions: what is the next place a user will visit? And how long will he stay in the current place? To tackle these issues, mobility and contextual patterns of a user using a probabilistic method are extracted and combined. The key idea in this method is based on an assumption that human mobility can be explained by multiple mobility patterns that depend on a subset of the contextual variables. Using smartphone data collected from 153 users, potential of this method in predicting human mobility is evaluated.

NextPlace (Scellato, Musolesi, 2011) is a prediction method which focuses on the temporal predictability of users when they visit their most important places. In this method, a time-location prediction method based on nonlinear time series analysis of the arrival and staying times of users in relevant places are considered. This method does not focus on the transitions between different locations: instead, it focuses on the estimation of the duration of a visit to a certain location and of the intervals between two subsequent visits. The simulation results show an overall prediction precision of up to 90% and a performance increment of at least 50% with respect to the state of the art.

The authors (Yohan et al., 2012) quantify the predictable regularity in human behavior in order to extract meaningful properties in humans' behavior and predict their duration of stay in next locations. To address this issue, mobility information of 10 users, with room-level accuracy in indoor as well as outdoor environments, is recorded every two minutes for an entire day. Then, location-dependent and location-independent models with several feature-aided schemes are evaluated. The experimental results show that a location-dependent predictor is better than a location-independent predictor for predicting temporal behavior of individual user.

5.3. Contact Prediction

Predicting future contacts of mobile users based on their mobility patterns as well as their social and connectivity properties can be utilized to improve the efficient of networking protocols in OppNets. For example, forecasting a user's future contacts in a routing algorithm such as a method in (Nguyen and Giordano, 2012) can help the algorithm decide if a message stored by a node should be further carried or forwarded, and to which intermediate node it should be forwarded in order to obtain the best possible delivery ratio and latency. In some cases, the location and time of a contact can also be predicted while forecasting contact properties. In this subsection, some well-known contact prediction methods are introduced.

The correlation between mobility features, friend relationships, and contact probability have been explored in some recent literature. The authors (Eagle et al., 2009) find a close relation between users' movement patterns and their friendship relationships. Analyzing relational dynamics of individuals, they show that people are more likely to meet their friends than strangers. Researchers (Cho et al., 2011) investigate the interrelationship between travel distance and social relationships and show that social relationships can explain about 10% to 30% of all human movement, while periodic behavior explains 50% to 70%.

The work (Jahanbakhsh et al., 2012) explore the problem of inferring the missing part of a contact graph from those partial contact graphs by computing similarities between neighbor sets of external nodes. The experimental results using real life mobility traces show that time-spatial based scores provide the most reliable results for predicting missing contacts among external nodes. Furthermore, it is concluded that combining social information with time-spatial information provides better performance results than using each of them independently.

Jyotish[2] (Vu, Do, 2011) is a context prediction method which aims to predict a user's future visiting location, stay duration and future contact. In order to construct such a predictive model, Jyotish includes a clustering algorithm to cluster Wi-Fi access point information into different locations. Then, a Naive Bayesian classifier is constructed to assign these locations to records in a Bluetooth trace which results in a fine granularity of people movement. Then, the fine grain movement trace is used to construct the predictive model to provide answers for the three prediction parameters. Evaluation over the Wi-Fi/Bluetooth trace collected by 50 participants shows that Wi-Fi access point information can be used to infer location while Bluetooth traces can infer contact. The joint Wi-Fi/Bluetooth trace thus can be used to study people movement with a considerably high accuracy.

## 6. Open Issues

In light of the many work on human mobility models for OppNets, focusing on its various aspects, there are still many research questions left without any proper answer. In this section, we go one step further by presenting some future research directions, which brings new visions into the horizon of human mobility research.

*Modeling Large-scale Human Mobility:* collecting mobility data sets and modeling human movement in large scale have many applications such as in communication network validation and epidemiological studies. For example, it can be used to analyze the spread of infectious diseases (Meloni et al., 2011). Nevertheless, there are several challenges and problems for the researchers to collect large scale mobility traces such as high experimental cost, lack of motivation applications, business models and organizations (Hui et al., 2010). Table 2 also verify that the majority of the available traces are collected in small scals and mobility on a wider scale is not considered sufficiently. Consequently, we believe that future work on human mobility modeling should also take into account the behavior of populations at large scales.

Recent studies in this area mainly have strived to either generate large scale mobility models or analysis the characteristics of mobility models using large scale data sets. The authors (Mu et al., 2013) present a mobility map construction scheme for large-scale Wi-Fi mobility tracking in indoor areas to acquire users' normal daily activities. Similarly, WHERE (Isaacman et al., 2012) takes as input spatial and temporal probability distributions drawn from empirical data, such as Call Detail Records (CDRs), and produces synthetic CDRs for a synthetic population. In another work, (Stopczynski et al., 2013) propose a mobile sensing technique for collecting spatio-temporal and scial data about human mobility using the capabilities of Bluetooth capable smartphones carried by participants. This technique is utilized in a large music festival with 130000 participants where a small subset of participants installed Bluetooth sensing apps on their personal smartphones. To characterize large scale traces, (Hasan et al., 2013) explore human activity data in urban scale using LBSN data sets such as the Foursquare. Specifically, aggregate activity patterns of humans by finding the distributions of different activity categories over a city geography are characterized. Furthermore, individual activity patterns by finding the timing distribution of visiting different places are determined.

*Data Forwarding based on Mobility Prediction:* Data routing and dissemination in OppNets is a crucial issue since users contacts with each other opportunistically and there is not an end-to-end path between the source and destination of a message. To address this issue, several forwarding protocols have been proposed which take advantage of mobility prediction to forecast future movement of users and improve the routing performance. However, most of existing protocols such as PROPHET (Lindgren et al., 2003) focus on the contact prediction of mobility without considering the spatial and temporal aspects of the contact.

Exploiting users' context information such as their spatial and temporal information, in addition to contact prediction, can be used to select the best relay nodes which results in both a better resource usage and higher delivery ratio. CiPRO (Nguyen and Giordano, 2012) is a prediction-based routing algorithm which considers both spatial and temporal dimensions of a contact, so that the source device knows when and where to start the routing process to minimize the network delay and overhead. Similarly, predict and relay (PER) (Quan, Cardei, 2012) considers the time of the contact and determines the probability distribution of future contact times and choose a proper next-hop in order to improve the end-to-end delivery probability. SPRINT (Ciobanu et al., 2013) uses online social information of users to increase the probability of successful message delivery. As future investigations, other unexplored features of users such as their mobility patters based on weekly schedule can be predicted to improve the performance of the forwarding algorithms.

---

[2] In Sanskrit, Jyotish (Ji-o-tish) is the person who predicts future events.

## 7. Conclusion

Recent trend for human mobility in opportunistic networks (OppNets) is looking at individual and collective behaviors of mobile carriers, inferred from the social nature of human motion. Consequently, various aspects of human mobility behavior have been explored in the literature. In this paper, we presented a survey on human mobility from three major aspects: human mobility characteristics, human mobility models, and human mobility prediction methods. First, we discussed spatio-temporal and connectivity features of human mobility. Second, real world movement traces which are captured using Bluetooth/Wi-Fi enabled devices or on-line location-based social networks were outlined. Furthermore, a comprehensive review on simulation-based mobility models was presented and their most important characteristics were featured. Third, recently proposed human mobility prediction methods as a young and exciting research area which has many applications in designing protocols in OppNets were discussed. Finally, some major open research issues were explored and future research directions were outlined. We hope that this effort instigate future research on this critical topic encouraging mobile application and system designers to develop appealing human mobility models.

**Acknowledgements**

This research is sponsored in part by the National Natur al Science Foundation of China and the Fundamental Research Funds for the Central Universities (contract/grant number: No.61173179 and No.61202441). This research is also sponsored in part by the Fundamental Research Funds for the Central Universities (No. DUT13JS10).